\documentclass[12pt]{article}
\pdfoutput=1
\usepackage{hyperref}
\usepackage{graphicx}
\usepackage{graphics}
\usepackage{dsfont}
\usepackage{epsfig}
\usepackage{amsmath,amssymb,amsthm,amscd}
\usepackage{simpler-wick}

\newcommand{\Tr}{\textrm{Tr}}

\def\ket{\rangle}
\def\bra{\langle}

\newcommand{\be}{\begin{equation}}
\newcommand{\ee}{\end{equation}}
\newcommand{\ba}{\begin{aligned}}
\newcommand{\ea}{\end{aligned}}

\numberwithin{equation}{section}

\begin{document}
\begin{titlepage}

\rightline{USTC-ICTS/PCFT-25-53}

\vskip 3 cm

\centerline{\Large 
\bf  
Gibbons-Hawking Entropy and BMN Strings   }

\vskip 0.5 cm

\renewcommand{\thefootnote}{\fnsymbol{footnote}}
\vskip 30pt \centerline{ {\large \rm 
Min-xin Huang\footnote{minxin@ustc.edu.cn}  
} } \vskip .5cm  \vskip 20pt 

\begin{center}
{Interdisciplinary Center for Theoretical Study,  \\ \vskip 0.1cm  University of Science and Technology of China,  Hefei, Anhui 230026, China} 
 \\ \vskip 0.3 cm
{Peng Huanwu Center for Fundamental Theory,  \\ \vskip 0.1cm  Hefei, Anhui 230026, China} 
\end{center}

\setcounter{footnote}{0}
\renewcommand{\thefootnote}{\arabic{footnote}}
\vskip 40pt
\begin{abstract}

We provide some up-to-date discussions related to cosmological event horizon and entropy of our universe, then introduce an intriguing idea that there may be a universal finite upper bound for entropy accessible to an observer in consistent theories of quantum gravity.  We argue that the Berenstein-Maldacena-Nastase (BMN) strings provide a test of the idea. More speculatively, in an optimistic scenario,  this also provides a possible  estimate of the cosmological constant.

\end{abstract}

\end{titlepage}
\vfill \eject


\newpage

\baselineskip=16pt

\tableofcontents

\section{Introduction}

It is well known that our universe is dominated by dark energy, a kind of energy with negative pressure, which causes the observed accelerated expansion of the universe. The nature of dark energy is not entirely clear, but one possibility is that it is a positive cosmological constant, which is the assumption of the $\Lambda$ cold dark matter ($\Lambda$CDM) model, sometimes known as the standard model of cosmology. Although the recent DESI experiment results provide some indications that dark energy may not be a simple cosmological constant but might evolve over time, to our knowledge, a cosmological constant is still the simplest and most likely possibility at the moment. In this paper, we will follow this assumption. 

The cosmological constant $\Lambda$ comes from vacuum energy in quantum field theory. In generic scenarios, this calculation is divergent so naturally it should be of the order of the UV cut-off Planck energy. However, the observed value is about $10^{-122}$ of the naive expectation. This is the famous cosmological constant problem. We should note that before the discovery of accelerated expansion by supernovae experiments in the late 1990s, it was commonly believed that $\Lambda$ is probably zero and many mechanisms were proposed to explain this exact cancellation.  A small but non-zero value seems more difficult to explain. However this also provides an opportunity that if we can find a natural way to estimate the right order of magnitude  of the cosmological constant, the arguments would be more quantitative and convincing.

There have been many proposed solutions of the cosmological constant problem, but there is no universally accepted consensus at the moment. We will consider this problem from the perspective of entropy.  In this paper we put aside the possibility of $\Lambda$  being zero or negative, and focus on the case of asymptotic de Sitter space with a positive $\Lambda$. In this case, there is a cosmological event horizon where a light signal sent by an observer cannot reach even after an infinite travel time, due to the accelerated expansion of the space. In the 1970s, Gibbons and Hawking  proposed that similar to the Bekenstein-Hawking black hole entropy \cite{Bekenstein:1973ur, Hawking:1975vcx}, the cosmological horizon also measures the entropy of degrees of freedom beyond the horizon \cite{Gibbons:1977mu}.  The Gibbons-Hawking entropy is proportional to the area of the cosmological horizon, and is inversely proportional to the cosmological constant $\Lambda$. So for our universe, this entropy is about $10^{122}$. (In this paper we use the convention $\hbar=c=k_B=1$ and entropy is a dimensionless quantity.) If one can find a natural way to estimate the entropy of our universe, we can then also obtain an estimate for the cosmological constant.   

Entropy has indeed played a significant role in our understandings of quantum gravity.  One of the breakthroughs is the exact counting of the microstates of some supersymmetric black holes in string theory, first demonstrated by Strominger and Vafa \cite{Strominger:1996sh}.  Holographic entanglement entropy has led to significant progress in the famous black hole information paradox, see the review \cite{Almheiri:2020cfm}.  More recently, a seminal approach for dealing with general black holes in general relativity using path integral was proposed in \cite{Balasubramanian:2022gmo}.  

The origin of Gibbons-Hawking entropy is more mysterious. We will not attempt to provide a microscopic counting of Gibbons-Hawking entropy of a general de Sitter space. Instead, we propose the possibility of the existence of a universal finite upper bound on entropy accessible to an observer in quantum gravity under some reasonable restrictions. We will explain more details that although the Hilbert space can be infinite dimensional, there is no robust counterexample where entropy can become arbitrarily large in a reasonable physical process, taking into account the full effects of quantum gravity. Expanding upon our previous brief discussion in \cite{Huang:2023dyw}, we will show that it is possible to probe such a hypothetical bound in an idealized situation of type IIB pp-wave background using the  Berenstein-Maldacena-Nastase (BMN)  strings \cite{Berenstein:2002jq}. If our universe is a typical universe, its  Gibbons-Hawking entropy  should not be  too much smaller than the hypothetical bound. So in an optimistic scenario, this provides a possible estimate of the cosmological constant.

\section{Entropy of our universe}

\subsection{Gibbons-Hawking entropy}

We focus on late time cosmology where dark energy and matter dominate, so for simplicity we can neglect the radiation and curvature contributions. The flat  Friedmann-Lemaitre-Robertson-Walker metric is 
\be
ds^2 = dt^2 - a^2(t) (dx^2 +dy^2 +dz^2) , 
\ee
where $a(t)$ is the scale factor, which can be rescaled and does not appear in physically measured quantities, so it is convenient to set $a(t_0)=1$ at our present time $t_0$. Assuming the dark energy is a cosmological constant $\Lambda$,  the Friedman equation is 
\be \label{Hubble279} 
H(t)  = \frac{H_0}{a(t)^{\frac{3}{2}}} [ \Omega_m + \Omega_{\Lambda} a(t)^3 ]^{\frac{1}{2}} ,
\ee 
where $H =\frac{\dot{a}}{a}$ is the Hubble parameter, and $H_0$ is the Hubble parameter at current time $t_0$.  The current dark energy and matter components are about $\Omega_{\Lambda}=0.69$ and  $\Omega_m=0.31$ and the radiation component is negligible $\Omega_R\sim 10^{-4}$. Equation (\ref{Hubble279}) actually has an analytic solution for the scale factor in terms of simple functions but will not be needed here. 

There are several concepts of horizon in cosmology which are nicely explained  e.g. in the textbook \cite{Weinberg:2008zzc}. The Hubble horizon is simply the place where the speed of the cosmic expansion is the speed of light, and has a radius  $H_0^{-1}$.  The particle horizon concerns the past. The radius of the particle horizon is the distance that a light signal has traveled from the beginning of the time $t=0$. During early universe inflation epoch, the particle horizon expanded rapidly, solving the famous horizon problem. Neglecting the early universe epoch, using the Friedman equation (\ref{Hubble279}), we can calculate the distance that a light signal has traveled since radiation became negligible with the integral 
 \be \label{particle282}
 d_p =  \int_0^{t_0} \frac{d t} {a} =  H_0^{-1} \int_0^1 \frac{ d a} {a^{\frac{1}{2}} \sqrt{ \Omega_m + \Omega_{\Lambda} a^3}  }  =3.26  H_0^{-1}. 
 \ee
 Because of the expansion of the universe, this is somewhat bigger than the Hubble radius. This is roughly the radius of our observable universe.   
 
 In our context, we are mostly interested in the cosmological event horizon (CEH).  This horizon concerns the future. Its radius is the farthest distance that a light signal emitted now can travel, can be computed by the integral 
  \be  \label{event282}
  d_e =    \int_{t_0}^{\infty} \frac{d t} {a}  . 
 \ee
 Unlike the particle horizon, there may not be an event horizon. Only if the integral converges, then there is an event horizon. This is roughly equivalent to an accelerating expansion $\ddot{a}>0$. For example , if the scale factor grows like a power law $a\sim t^c$ with $c>0$, then $\ddot{a}>0$ is equivalent to $c>1$ and the convergence of  the integral (\ref{event282}). If the scale factor grows exponentially as in the case of a de Sitter space, then the integral (\ref{event282}) certainly converges more rapidly.  On the other hand, for radiation or matter dominated universe there is no event horizon.
 
 More generally, if  $\ddot{a}>0$, then we can estimate for $t>t_0$ that $\dot{a}(t) > \dot{a}(t_0)$ and $a(t)>   \dot{a}(t_0) (t -t_0) +a(t_0)$, so the integral (\ref{event282}) is generically convergent. However, this is not a sufficient condition. One can construct some contrived examples where the scale factor grows barely linearly at late time such that the integral (\ref{event282}) is divergent. One example is $a(t) \sim t \log^{\alpha}(t)$. For the range $0< \alpha <1$, this scale factor function satisfies $\ddot{a}>0$ at late time but the integral (\ref{event282}) is divergent. There seems no natural physical realization for such contrived scale factor functions. 

For our universe we can compute the current radius of CEH with  the integral 
  \be  \label{devent}
  d_e =    H_0^{-1} \int_1^{\infty} \frac{ d a} {a^{\frac{1}{2}} \sqrt{ \Omega_m + \Omega_{\Lambda} a^3}  }  =1.15H_0^{-1}.  
 \ee
For a 4D pure  de Sitter space with the cosmological constant  $\Lambda>0$, the Hubble parameter $H_{\Lambda} = \sqrt{\frac{\Lambda}{3}}$ is independent of time, and the radius of CEH is simply $H_{\Lambda}^{-1} $. The radius (\ref{devent}) is close to the current Hubble radius $H_0^{-1}$ because our universe is already 69\% dominated by dark energy. In the far future, our universe will be completely dominated by the cosmological constant and asymptote to a de Sitter space, and the radius of CEH becomes slightly larger $H_{\Lambda}^{-1}  = H_0^{-1}\Omega_{\Lambda}^{-\frac{1}{2}} = 1.20 H_0^{-1}$. 

In the unit of Planck length, the current observed Hubble radius is $H_0^{-1}\approx 8.5\times 10^{60} l_p$, so we can calculate the value of the cosmological constant and the far-future Gibbons-Hawking entropy as one quarter of the CEH area  
\be
\Lambda = 3 H_0^2 \Omega_\Lambda \approx 2.9 \times 10^{-122} l_p^{-2}, ~~~~  S = \frac{\pi  H_0 ^{-2} } {\Omega_\Lambda l_p^2}  \approx 3.3 \times 10^{122}. 
\ee

Before the late 1990s, some occasional pioneering works did seriously consider a non-zero $\Lambda$ for our universe. Notably, Weinberg's anthropic argument \cite{Weinberg:1987dv} essentially predicted the cosmological coincidence, i.e., the current energy densities of $\Lambda$ and matter are comparable $\Omega_{\Lambda}\sim \Omega_{m}$. If $\Omega_{\Lambda}$ is order one, then    we see that the order of magnitude of $\Lambda$ is determined by the observed Hubble constant.  It is a much more difficult problem to provide a purely theoretical derivation of this absolute magnitude $10^{-122}$ in Planck units.

We provide some more general discussions about Gibbons-Hawking entropy. We note that the calculation  is not affected by possible extra compact dimensions and their sizes. Suppose we compactify from $d$ dimensions to 4 dimensions, then the length scales of gravity are related $l_d^{d-2} = V_{d-4} l_p^2$, where $ V_{d-4}$ is the compactification volume. The entropy is the same for the $d$-dimensional calculation $S =\frac{  V_{d-4} A}{4 l_d^{d-2}}= \frac{ A}{4 l_p^2}  $, where $A$ is the 4D horizon area.

The derivation of the temperatures of a black hole and de Sitter space are similar. One can Wick rotate to Euclidean time, and then expand around the horizon. Demanding no conical singularity determines the periodicity of the Euclidean time, or the inverse of the temperature. For black hole, one can derive the entropy using the first law of thermodynamics $dE=TdS$. However for de Sitter space, this derivation encounters a  subtlety of using negative energy, since adding mass to an empty de Sitter space decreases the horizon area. See e.g. the review paper \cite{Spradlin:2001pw} for more discussions. 

One can also compute the entropy from the partition function using the thermodynamic relation 
\be  \label{entropy296}
S = \log(Z) + T\partial_T \log(Z) ,  
\ee 
where the partition function can be computed from the Euclidean gravity action $Z=e^{-S_E}$.  For the black hole case, the Euclidean action $S_E$ is positive and both terms in (\ref{entropy296}) contribute to give the correct entropy. However, for the de Sitter space, the Euclidean action $S_E$ is negative, and one approach is to simply neglect the second term in (\ref{entropy296}) and use $S\simeq \log(Z)$ to get the correct Gibbons-Hawking entropy, plausibly because the de Sitter temperature is determined by the cosmological constant, a fixed quantity that we do not vary here. See, e.g., a recent paper \cite{Maldacena:2024spf}.  The situation is somewhat similar to our earlier paper \cite{Huang:2023dyw}, where  the first term in the formula  (\ref{entropy296})  dominates in the high temperature limit in the thermal entropy of a quantum system. Furthermore, the one-loop correction to the de Sitter action may be imaginary, an issue recently addressed also in \cite{Maldacena:2024spf}. In any case, despite the subtleties, the entropic interpretation of an event horizon is now well supported by the general principles of holography, and shall be a valid foundation for our further discussions.

\subsection{Other entropies} 

There have been some estimates for the entropies of various components of the observable universe in astrophysics, see e.g. \cite{Egan:2009yy, Profumo:2024hnn}. In generic scenarios, the dominant contribution to entropy comes from supermassive black holes (SMBH), which reside at the center of most galaxies, including our own Milky Way.  A typical SMBH may have a mass of  $10^8$ times that of the Sun and an entropy of $10^{93}$. It is estimated that there are no more than $10^{12}$ galaxies in the observable universe. So the total entropy of SMBHs is about $10^{105}$.  How SMBHs formed so early and grew so huge is still one of the big open questions in astrophysics. 

For comparison, the entropy of CMB photons is about $10^{88}$, and the entropy of dark matter is about $10^{79} - 10^{81}$ in most models. Although dark matter accounts for about 25\% energy of our universe, the entropy contribution is small because in most models it is very cold and weakly interacting. It is perhaps not too surprising that black holes contribute the most since they are a kind of maximal entropy objects. For example, the entropy of the Sun is about $10^{58}$, while it may increase dramatically to about $10^{77}$ if the same mass collapses into a black hole. 

The Gibbons-Hawking entropy is usually regarded as coming from degrees of freedom beyond the future cosmological event horizon (CEH).
Because the radius of our observable universe (\ref{particle282}) is larger than the radius of CEH (\ref{devent}),  a significant fraction of galaxies in the observable universe are actually beyond CEH.  In standard cosmology, a distant galaxy just goes with the Hubble flow, i.e.  its comoving distance from us is fixed while its physical distance is proportional to the scale factor. The worldline of a galaxy intersects the past light cone of our present event if the galaxy is in the observable universe, while its worldline may lie outside our future light cone if its current physical distance from us is larger than the radius of CEH.  Equivalently, we can still receive old light from such a galaxy, but a signal emitted by us today will never reach the galaxy.   From our perspective, a galaxy will appear to freeze at an event point where its worldline intersects with the past light cone of our future infinity and eventually fade away due to increasing cosmological redshift.

In the far future, all matter will move beyond the CEH and the entropy inside the CEH will decrease. Meanwhile, as we mentioned, the Gibbons-Hawking entropy will slightly increase, more than compensating for the loss, so that the total entropy always increases, consistent with the second law of thermodynamics.  From this perspective, the total entropy of localized objects inside the current CEH must be smaller than the far-future Gibbons-Hawking entropy minus the current one, although the volume scales like the cube of the radius.  It should be emphasized that the far-future empty de Sitter gives the maximal entropy. Localized excitations are entropy-deficit states and should not be naively added to the Gibbons-Hawking entropy of the empty de Sitter, but rather to that of the actual  geometry containing the excitations. This is sometimes called the generalized entropy which obeys the second law of thermodynamics. 

We have provided an explicit discussion  that for our universe, the total entropy of various components is indeed negligible compared to the Gibbons-Hawking entropy of about $10^{122}$.  Due to the cosmological coincidence,  the matter energy density is $\rho_m \sim \frac{3H_0^2}{8\pi G}$, so the total mass inside our current CEH is $M= \frac{4\pi}{3} \rho_m d_e^3 \sim \frac{1}{2G H_0}\sim 10^{60} m_P$, where $m_P$ is the Planck mass. If all matter inside the CEH were turned into a  single  Schwarzschild black hole of the same mass, it would have an entropy comparable to the Gibbons-Hawking entropy. Since a black hole is a maximal entropy object, this is another argument why the actual localized objects inside our CEH should have a much smaller and negligible entropy. In any case, for our purpose, the Gibbons-Hawking entropy is a very good approximation for the entropy of our universe.

\section{A universal finite upper bound for entropy?}

The entropy of our universe is finite ($\sim 10^{122}$) and will not increase much in the far future if dark energy is a cosmological constant  as we have assumed here. Is this a hint for a universal finite upper bound for entropy accessible to an observer in consistent theories of quantum gravity? Of course in practice it is impossible to construct a physical system whose entropy is much larger than that of our own universe. We argue that this may not be possible even in principle. Bekenstein has famously conjectured an entropy bound for a physical system with a given size \cite{Bekenstein:1980jp}. This was applied to cosmology and further generalized in, e.g., \cite{Fischler:1998st, Bousso:1999xy}. Here we conjecture the existence of a fundamental finite entropy bound independent of the system size or any other parameters, under certain conditions which will be explained shortly. We should emphasize that we are making a conjecture about quantum gravity in general, not just for de Sitter space, which will not be the focus of this section. 

In classical physics, the entropy is only defined up to a constant by the first law of thermodynamics. Before the advent of quantum mechanics, the statistical interpretation requires artificially dividing the continuous phase space into discrete cells, also introducing an ambiguous constant in entropy. In order to have a true definition of entropy without any ambiguous constant, we need to consider quantum theory and use von Neumann entropy. We will consider a density matrix  $\hat{\rho}$ which is a Hermitian, positive semidefinite, trace class operator on a Hilbert space so that we can normalize $\Tr(\hat{\rho})=1$.  The von Neumann entropy is defined by 
\be
S = -\Tr (\hat{\rho}\log(\hat{\rho})) = - \sum_i p_i\log(p_i), 
\ee
where $p_i$'s are the eigenvalues of $\hat{\rho}$ and form a probability distribution. In information science, this is also the well known formula for Shannon entropy. Some general properties of entropy were reviewed in \cite{Wehrl:1978zz, Witten:2018zva}.

In quantum physics, one often considers states with continuous quantum number, such as position or momentum eigenstates. These states are not normalizable but are still quite useful, e.g. in scattering problems where the particles go to infinity. They do not in a mathematical rigorous sense belong to a Hilbert space, which is assumed to be separable or have a countable basis.  Instead, they belong to a generalized space called the rigged Hilbert space. One can still define an entropy analogously using the probability density in the continuous phase space. However, the definition is only a semiclassical  approximation. In particular, the entropy  can be negative if the probability density is not too smooth \cite{Wehrl:1978zz}. In this paper we shall not consider further the case of continuous spectrum. 

Recently there have been many research activities relating to von Neumann algebras in string theory community, see the review papers \cite{Witten:2018zxz, Sorce:2023fdx, Liu:2025krl}. There are three types of von Neumann algebras. Type III algebra appears in the study of algebraic quantum field theory. In this case, trace and entropy are not defined, and only a relative entropy can be defined. In type II  algebra, one can define a renormalized trace and the entropy is defined up to an ambiguous constant. Here the renormalized trace is not the usual trace over Hilbert space as most operators in the algebra are not trace class. In the seminal papers \cite{Chandrasekaran:2022cip, Witten:2023xze}, the identity operator which has maximal entropy in type II$_1$ algebra is proposed to describe the empty de Sitter space. Other operators have smaller entropy but there is no lower bound. In order to test our conjecture, we need to have a definition of an absolute non-negative entropy. So in this paper in the followings we focus on the density matrices in the simplest type I algebra, which satisfy this criterion.  

In a finite dimensional Hilbert space, the maximal von Neumann entropy is the logarithm of the dimension, obtained by the properly normalized identity operator. It is usually assumed that the thermodynamic macroscopic entropy is this maximal entropy. For example in the microstate counting for black holes, one only needs to compute the dimension of the relevant Hilbert space. However, if the dimension of the Hilbert space is infinite or astronomically large, it is not clear whether this theoretical maximal entropy can always be obtained in a reasonable physical process.  

Banks has proposed that de Sitter space is described by a finite dimensional Hilbert space \cite{Banks:2000fe} \footnote{This idea was introduced independently in a talk ``Taking de Sitter Seriously" by W. Fischler at the Festschrift for G. West's 60th birthday in New Mexico in 2000.}. See also some early related discussions in \cite{Balasubramanian:2001rb, Witten:2001kn}. However, as mentioned in \cite{Witten:2001kn}, for our universe, the dimension of such Hilbert space would be astronomically large, and at present we are still not aware of a potential theoretical method to derive it. In this paper in the followings we will instead focus on infinite dimensional Hilbert spaces and explore potential methods to derive large but finite entropy. 

In an infinite dimensional Hilbert space, there is not a unique density matrix with maximal entropy, because the identity operator is not trace class. To get infinite von Neumann entropy, as mentioned in \cite{Wehrl:1978zz, Huang:2023dyw}, one can consider a rather contrived probability distribution with the asymptotic behavior $p_n \sim \frac{1}{n\log^\alpha(n)}$ where $1<\alpha \leq 2$. More general probability distributions with infinite entropy were discussed in \cite{Baccetti:2013ctj}. They are rather contrived but they can get arbitrarily close to any density matrix with finite entropy. In this sense the density matrices with finite entropy form a ``meager" subset in the usual topology \cite{Wehrl:1978zz}. However, a meager set does not necessarily mean that it is negligibly small. There are examples of meager sets that have positive Lebesgue measure.  In any case, we are not aware of any reasonable physical situation where such contrived probability distributions can appear. For example, in a quantum system with an infinite dimensional Hilbert space, such as a harmonic oscillator, the thermal entropy is indeed finite at a finite temperature, and tends to infinity in the infinite temperature limit \cite{Huang:2023dyw}. So a more relevant question is whether the entropy is bounded or unbounded in physically allowed range of parameters. 

For the quantum systems without gravity, the thermal entropy is unbounded \cite{Huang:2023dyw}. We would like to argue that when the full effects of quantum gravity are taken into account, the entropy may be bounded by a universal finite value in any reasonable physical process. Indeed, it is commonly expected that quantum gravity effects will be important when the temperature is sufficiently large. For the Calabi-Yau models in \cite{Huang:2023dyw}, the entropy grows much slower than that of conventional models at high temperature, hinting at a possible manifestation of quantum gravity effects in the right direction. 

We now describe a physical process to generate entropy. We start with a pure state, which has zero von Neumann entropy. The initial state can then undergo a quantum evolution which we assume to be unitary. A unitary transformation does not change the von Neumann entropy. In order to have a finite entropy, we allow for an external observer to make a measurement.  In the usual interpretation of quantum measurements, the quantum state is projected into an orthogonal basis of the Hilbert space, and we have a probability distribution for the outcomes of the measurement. Suppose the external observer does not reveal the result of the measurement, then we have a mixed state with finite entropy. This is the usual decoherence process where the environment plays the role of an external observer.  This can be also understood as an entanglement entropy between the quantum state and the observer.

Some measurement processes do not seem to be affected by quantum gravity consideration, but can indeed produce arbitrarily large or even infinite entropy. In order for our proposal to be valid, we need to rule out these processes as ``unreasonable".  Firstly, if the observer is allowed to choose any orthogonal basis, then one can actually produce a mixed state with any probability distribution $\sum_{n=0}^{\infty} p_n=1$. Suppose we have a pure normalized state $|\psi \ket$, we can constructed an orthonormal basis inductively. We can choose a normalized state $|\psi_0\ket$ such that $|\bra \psi | \psi_0\ket|^2 =p_0$. Then in the subspace orthogonal to $|\psi_0\ket$, we further choose $|\psi_1\ket$ such that $|\bra \psi | \psi_1\ket|^2 =p_1$, and so on. If the observer makes a measurement with respect to the basis $|\psi_n\ket, n=0,1,2, \cdots $, we get a mixed state $\sum_{n=0}^{\infty} p_n  |\psi_n\ket \bra \psi_n |$ which could have arbitrary or infinite entropy. Similarly, if we can choose an arbitrary initial state, then it is  also easy to get arbitrary entropy after the measurement. In order to avoid this, we require the observer only makes measurements in some predetermined natural bases. Furthermore, the initial state is also required to be a basis state or a simple superposition, which can then undergo unitary evolution without control from the observer. 

Secondly, we show that if the observer is allowed to make an unlimited number of measurements, each time without memory of the previous measurement results, then the entropy is likely to become arbitrarily large. So we require the observer to make only one or at most a small number of measurements. In our macroscopic world we encounter decoherence frequently. However, our universe as a whole may well be in a pure state if there is no measurement from an external observer. This is indeed an often assumption in many studies in quantum cosmology. It seems reasonable to limit the number of measurements. 

To show this, we consider a Hilbert space with an orthonormal basis $|n\ket, n\in \mathbb{Z}$. We consider a unitary transformation $\hat{U}$ that the transition probability between two basis states only depends on their difference
\be 
p_{m-n} = |\bra m | \hat{U} |n \ket |^2 ,  
\ee
where $p_n$'s form a probability distribution. A simple example is the symmetric random walk, where one moves left or right by one unit  with equal probability, i.e. $p_1=p_{-1}=\frac{1}{2}$ and all other $p_n$'s vanish. Suppose we start with the state $|0\ket$ and perform the unitarity transformation and an external measurement in the basis $|n\ket, n\in \mathbb{Z}$, then we have a mixed state $\sum_{n=-\infty}^{\infty} p_n |n\ket \bra n|$, which can be equivalently represented as a random variable taking values in integers with the prescribed probability. If we repeat this procedure $N$ times, we get a mixed state represented by the sum of $N$ such random variables. When $N$ is large, according to the Central Limit Theorem, the probability is well approximated by a normal distribution whose variance is proportional to $N$. We can compute the entropy with the continuous normal distribution $f(x) = \frac{1}{\sqrt{2\pi} \sigma}e^ {-\frac{x^2}{2\sigma^2}}$ as a good approximation if $\sigma$ is large
\be
S_\sigma = - \int_{-\infty}^{\infty} f(x) \log f(x) dx  =\log(\sqrt{2\pi} \sigma )+\frac{1}{2}, 
\ee 
which is clearly unbounded as $\sigma\rightarrow \infty$. 

Ruling out these unreasonable scenarios, we now consider the entropy of BMN strings. In general we do not have a full theory of quantum gravity, so it seems impossible to test our conjecture. However, in a very special type IIB string pp-wave background with infinite spacetime curvature and Ramond-Ramond flux, we conjectured that we had a full theory of quantum gravity described by tensionless BMN closed strings \cite{Huang:2019lso, Huang:2019uue, Du:2021dml, Du:2021spv}. For simplicity we consider the case of two string modes and denote the  BMN strings by the complete orthonormal basis $|n\ket, n\in \mathbb{Z}$. By holographic correspondence, the BMN two-point function in free $\mathcal{N}=4$ $SU(N)$ super Yang-Mills theory including all genus contributions is identified with a probability or the norm square of the unitary transition amplitude 
\be 
p_{m,n}(g) = |\bra m | \hat{U}(g) |n \ket |^2,
\ee 
where $g=\frac{J^2}{N}$ is the  non-negative genus counting string coupling constant from the BMN double scaling limit with large R-charge $J\sim  \sqrt{N} \sim \infty$.  Because of the convergence of the genus expansion, we can in principle compute the probability $p_{m,n}(g)$ to arbitrary precision for any string coupling $g$. Following the reasonable process we described earlier with one measurement, we have the entropy of BMN strings \cite{Huang:2019uue} for an initial state $|m\ket$ at a coupling $g$ as 
\be 
S_m(g) = - \sum_{n=-\infty}^{\infty} p_{n, m}  \log ( p_{n, m} ). 
\ee
It was found that the entropy is indeed finite at a finite coupling $g$, but at the moment it is still not clear to us whether it is bounded or unbounded as $g\rightarrow \infty$. If it is bounded, this would be a non-trivial evidence for our conjecture of a universal finite bound. 
Since this is an extreme situation where classical gravity completely breaks down and infinitely many excited stringy states become degenerate, it may be a prime theoretical setting for testing an intrinsic quantum gravity bound. Of course this is far from a derivation, but in an optimistic scenario, we hope that the entropy supremum might not be too much smaller than the hypothetical bound, so might be comparable to the entropy of our universe. 

Perhaps an easier problem is to first understand the asymptotic behavior of $p_{n. m}(g)$ as $g\rightarrow \infty$. Here we prove a proposition  which provides a relevant general criterion for the (un)boundedness of entropy if the limit exists.  For simplicity we omit the initial state index $m$ and first prove an easier lemma. \textit{Lemma}: Suppose $\lim_{g\rightarrow \infty} p_n(g) = 0 $ uniformly in $n$, then the entropy $S(g) =-\sum_n p_n(g)\log p_n(g)$ tends to infinity as $g\rightarrow \infty$. This lemma applies for example to the thermal entropies in \cite{Huang:2023dyw}. The proof is not difficult. For any large natural number $N$,  we can find a $g_0$ such that when $g>g_0$, we have $p_n(g)< \frac{1}{N}$ for any $n$. We note that the function $f(x)=-x\log(x)$ is a concave function for $x\geq 0$  because its second derivative is always negative.  So the entropy is decreased if we move the $N$ largest $p_n(g)$'s to $\frac{1}{N}$ and the remaining ones to zero, keeping the sum fixed $\sum_np_n(g) =1$.  In this case the entropy is $\log(N)$, which tends to infinity as $N\rightarrow \infty$. 
 
 We note that the uniform condition in the convergence is necessary. For example, we consider an example similar to the one in the Appendix in \cite{Lieb:1973cp}. Suppose $p_n(g)= 1$ for $n=[g]$ and $p_n(g) = 0$ for other $n$'s. Then it is true that $\lim_{g\rightarrow \infty} p_n(g) = 0 $ for any $n$ but the convergence is not uniform. The entropy is always zero $S(g)=0$ for any $g$, so is certainly bounded. 
 
We now prove the stronger proposition. Suppose $\lim_{g\rightarrow \infty} p_n(g) = \tilde{p}_n$. It is easy to see $0\leq \sum_n \tilde{p}_n \leq 1$. The sum can be smaller than one because the infinite sum and limit does not necessarily commute.  \textit{Proposition}: Suppose $\lim_{g\rightarrow \infty} p_n(g) = \tilde{p}_n $ uniformly in $n$ and $\sum_n \tilde{p}_n<1$, then the entropy also tends to infinity as $g\rightarrow \infty$.  \textit{Proof}: We choose a fixed number $\delta$ in the range $0<\delta <1- \sum_n \tilde{p}_n$. For any large natural number $N$, since the sum $\sum_n \tilde{p}_n$ is convergent, only a finite number of $\tilde{p}_n$'s are larger than $\frac{\delta}{2N}$. We find a number $n_0$ such that when $|n|>n_0$, we have $\tilde{p}_n<\frac{\delta}{2N}$. Since the convergence is uniform in $n$, we can find a $g_0$ such that when $g>g_0$, we have $p_n(g)<\frac{\delta}{N}$ for $|n|>n_0$ and $\sum_{|n|\leq n_0} p_n(g) <1-\delta$. Similar to the proof of the lemma, the entropy is decreased if we move the $N$ largest $p_n(g)$'s in the range $|n|>n_0$ to $\frac{\delta}{N}$ while keeping the sum $\sum_{|n|> n_0} p_n(g) >\delta$ fixed. In this case the entropy is larger than $\delta \log({\frac{N}{\delta}})$, which tends to infinity as $N\rightarrow \infty$. 

Obviously the conditions cannot be further relaxed. If $\sum_n \tilde{p}_n= 1$, it is easy to construct examples such that the entropy is bounded as $g\rightarrow \infty$. In an infinite dimensional Hilbert space, entropy in general is not necessarily continuous and there are only a few known restricted continuity properties, e.g. the lower semicontinuity \cite{Wehrl:1978zz}. In mathematical analysis, Tannery's theorem  validates the interchange of infinite sum and limit if the sum is bounded by a fixed convergent sum. In physical situations, such as the thermal partition function at a finite temperature \cite{Huang:2023dyw}, or the entropy of BMN strings  at a finite coupling \cite{Huang:2019uue},  the  conditions for Tannery's theorem are easily verified to be satisfied, so the entropy is a continuous function. However the conditions may not be satisfied asymptotically,  as it happens in the infinite temperature limit. So in the case of $\sum_n \tilde{p}_n= 1$, regardless of whether $\tilde{S}:= -\sum_n \tilde{p}_n\log \tilde{p}_n$ is finite, in general it seems undetermined  whether $\lim_{g\rightarrow \infty} S(g)$ exists or is equal to $\tilde{S}$.

\section{Conclusion} 

If our conjecture is correct, it would be interesting to search for physical systems with an infinite dimensional Hilbert space where the entropy is bounded for the range of  all allowed  parameters. Regardless of whether it can be certain that the effects of quantum gravity are included, such systems may serve as useful models for computing the entropy of our universe.  On the other hand, we believe the entropy of BMN strings in \cite{Huang:2019uue} comes from a reasonable physical process where the full effects of quantum gravity have been  accounted for, so at least this case provides a definitive test of our conjecture.

One of the main problems of quantum gravity is the lack of experimental tests. To make progress on this issue, we may not need new experiments. Instead, one can try to compute from first principles the known parameters in physical models of particle physics or cosmology already determined by experiments. In string theory, although one can construct many phenomenological models qualitatively similar to the standard model of particle physics, at the moment there is no precise test due to the plethora of vacua in string landscape and the lack of a general non-perturbative formulation to select the vacuum. See, e.g., a recent review \cite{Marchesano:2024gul}. On the other hand, the cosmological constant is more likely to be a universal parameter independent of the details of low energy constructions, thus seems to be a more promising research target for experimental verification if we are fortunate. We hope to make more substantial progress in the future with our approach in this paper, in particular, settling the key question whether the entropy of BMN strings is bounded or unbounded.

\vspace{0.2in} {\leftline {\bf Acknowledgments}}
\nopagebreak

I would like to thank the organizers of the 2025 International Congress of Basic Science (ICBS) held  in Beijing in July 2025 and ``Workshop on QFT and Beyond" held in Southeast University in Nanjing in November 2025 for invitations and hospitalities where parts of the work are included in my presentations, and which provide the motivation for writing this paper.  I thank ChatGPT, Bao-ning Du, Albrecht Klemm for helpful discussions. This work is  supported by the National Natural Science Foundation of China Grants No. 12325502 and No. 12247103.  

\appendix

\addcontentsline{toc}{section}{References}

\bibliographystyle{utphys} 
\bibliography{ReferenceGHBMN}


\end{document}